# Error sources in heralded quantum Zeno gates

I.C. Nodurft and J.D. Franson

*Physics Department, University of Maryland Baltimore County, Baltimore, MD 21250 USA*

**Abstract:** Quantum logic gates for photonic qubits can be implemented using the quantum Zeno effect based on strong two-photon absorption. The fidelity of quantum Zeno gates of this kind may be substantially reduced by photon loss. Heralding on those outcomes in which both of the logical qubits emerge from the Zeno gate can increase the fidelity at the expense of a limited success rate (P.M. Leung et al., Phys. Rev. A **74**, 062325 (2006)). We analyze the performance of heralded quantum Zeno gates by solving Schrodinger's equation for a system of photons coupled to three-level atoms. This approach identifies several potential error sources that are not described by earlier models that assumed a fixed rate of single-photon loss and two-photon absorption.

## I. Introduction

Quantum logic gates for photonic qubits have a number of potential applications, including quantum computing and quantum repeaters. Quantum logic operations can be performed using linear optics techniques [1-5], but their inherent failure rate requires a large amount of overhead for quantum error correction. Quantum logic operations can also be performed using nonlinear effects in atoms or bulk materials [6-13], but in that case the performance is often limited by the strength of the nonlinear interaction or by photon loss. Here we consider an approach that combines quantum logic operations [14-29] based on the quantum Zeno effect [30-35] in a nonlinear medium with heralding techniques familiar from linear optics. This combined approach has several potential advantages over using either approach alone [15,16,18].

The quantum Zeno effect [30-35] uses frequent measurements to inhibit an undesirable process from occurring. Here, we will use the Zeno effect to inhibit a single photon in one wave guide from coupling into a second wave guide when another photon is already present there [14-18]. This can be accomplished by coupling the photons to an ensemble of three-level atoms with strong two-photon absorption. The quantum Zeno effect allows the implementation of a nonlinear phase gate (sign gate) at the two-photon level, which can be used to construct a controlled-NOT (CNOT) gate, for example [14-18].

One of the practical difficulties in implementing quantum Zeno gates in this way is the requirement for large two-photon absorption rates along with relatively small single-photon loss. It has previously been shown that the effects of photon loss can be greatly reduced by heralding on events in which both of the logical qubits exit from the Zeno gate. This can be done in a scalable way using fusion gates [15] or CNOT gates based on teleportation [16]. Photon loss codes can also perform a similar function [18].

Previous studies of heralded quantum Zeno gates were based on a model that assumed that photon loss could be described by fixed rates of single-photon loss and two-photon absorption [15,16,18]. Here we will use a more detailed model in which the photons in a Zeno gate interact with a set of three-level atoms to produce the necessary two-photon absorption. Rather than assuming fixed rates of photon loss, we will solve Schrodinger's equation for the combined system of photons and atoms. The results of our analysis show several types of error sources that are not included in the simple loss model, including errors in the controlled phase shift produced by the Zeno gate and loss due to nonadiabatic coupling. These additional error sources make it more difficult to correct the output of a Zeno gate using distillation [15,16] or photon loss codes [18].

We begin in Section II by reviewing the way in which the quantum Zeno effect can be used to implement quantum logic gates for photonic qubits. Section III describes the modelling and simulation of a system of coupled wave guides and three-level atoms that could be used to implement a Zeno nonlinear phase gate. The effects of limited two-photon absorption are discussed in section IV. The use of heralding to reduce the effects of single-photon loss and nonadiabatic wave guide coupling are described in sections V and VI. Experimental considerations are considered in section VII. A summary and conclusions are provided in Section VIII.

## II. Zeno effect and quantum logic gates

The quantum Zeno effect can be used to inhibit transitions from an initial state $\left| \psi_0 \right\rangle$ to an undesired final state $\left| \psi_F \right\rangle$, such as an error state, by making frequent measurements to determine whether or not the transition has occurred. For a sufficiently short time interval $\Delta t$ between the measurements, the probability amplitude for a transition will be proportional to $\Delta t$ while the probability itself will be proportional to $\Delta t^2$. As a result, a measurement after a short time interval $\Delta t$ will have a high probability of collapsing the system back into $|\psi_0\rangle$ with no remaining amplitude for $|\psi_F\rangle$. Our goal is to use heralding techniques to reduce the residual errors that are on the order of $\Delta t^2$. The output of a Zeno gate can also be corrected using other techniques, such as distillation or photon loss codes [15,16,18,21].

It is well known that an actual measurement is not required and that a strong interaction between the error state $|\psi_F\rangle$ and the environment is sufficient to implement the quantum Zeno effect [31]. Roughly speaking, information left in the environment could be used to determine whether or not the transition has occurred, and this will inhibit the transition regardless of whether or not any actual measurement is made. Here we use strong two-photon absorption to inhibit the growth of a state in which there are two or more photons in a wave guide. It is important to realize that no two-photon absorption actually occurs in the limit of a sufficiently strong two-photon absorption coefficient.

These effects can be used to implement a controlled phase gate (sign gate) as illustrated in Fig. 1 [14]. Two wave guides are gradually brought together in such a way that a photon initially present in the upper wave guide will be coupled into the lower wave guide by means of their evanescent fields. Wave guide devices of this kind with adjustable coupling are commercially available. By adjusting the strength of the coupling and the length of the coupling region, it is possible to ensure that a photon initially present in one wave guide will be completely coupled into the other wave guide as illustrated in Fig. 1(a). It can be shown that a photon will pick up a phase shift of $\pi/2$ in coupling from one wave guide to the other in addition to its usual phase propagation. The same phase factor occurs in the Rabi flopping of atoms [37] and it is also analogous to the $\pi/2$ phase difference between the reflection and transmission coefficients of a beam splitter [38].

Two wave guides coupled together in this way form a linear device in the absence of any two-photon absorption. A second photon initially present in the lower wave guide would be coupled into the upper wave guide and receive a phase shift of $\pi/2$ as well. If two photons propagate into the device, one in each wave guide, the system would be subjected to a total linear phase shift of $\pi$ in the absence of any two-photon absorption.

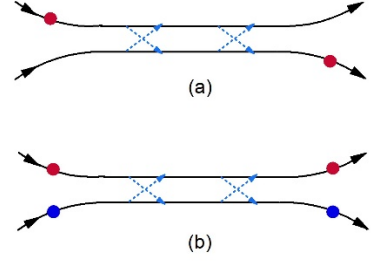

Fig. 1: Implementation of a controlled phase gate using the quantum Zeno effect [14]. Two wave guides are coupled together by their evanescent field as illustrated by the blue arrows, which allows photons to be transferred from one path to the other. The photons are coupled to a large number of three-level atoms (not shown) with strong two-photon absorption. (a) A single photon incident in the upper wave guide (red dot) will be coupled completely into the lower path, which produces a phase shift of $\pi/2$. A single photon present in the lower path (not shown) would also be coupled into the upper path with a phase shift of $\pi/2$. (b) When a single photon is incident in both paths, the Zeno effect due to strong two-photon absorption prevents either photon from being transferred to the other path. This eliminates the two phase shifts of $\pi/2$, giving a net controlled phase shift of $\pi$.

Now consider the situation in which both wave guides contain a strong two-photon absorbing material. Alternatively, this could be implemented using a collection of three-level atoms outside of the wave guides and coupled to the photons by means of their evanescent fields. If the two-photon absorption coefficient is sufficiently large, a single photon present in one of the wave guides will not be coupled into the other wave guide if a second photon is already present there, since the quantum Zeno effect would strongly suppress the growth of the two-photon probability amplitude [14]. Thus both photons will emerge from the same wave guide that they entered, as illustrated in Fig. 1(b).

Since the coupling of the photons from one wave guide to the other is inhibited in that case, the two phase shifts of $\pi/2$ do not occur when a photon is present in both wave guides. This corresponds to a nonlinear or controlled phase shift of $\pi$ that occurs for two incident photons as compared to the usual linear response. We assume a single-rail encoding in which the absence of a photon in a wave guide corresponds to a logical value of "0" while the presence of a photon corresponds to a logical "1". With that choice of qubits, the device shown in Fig. 1 produces a controlled phase shift of $\pi$ that can be used to implement a controlled-NOT gate, for example.

Quantum logic operations can also be implemented using the photon blockade effect that occurs when an atom is strongly coupled to a low-loss cavity [9]. In that case, a second photon is prevented from entering the cavity if another photon is already inside. Photon blockade is somewhat similar to the behavior seen in Fig. 1, with the main difference being that a quantum Zeno gate uses a dissipative effect whereas the photon blockade approach attempts to minimize any dissipation.

One of the main limitations of a quantum Zeno gate is that the coupling from one wave guide to the other will not be completely inhibited if the two-photon absorption is too weak. Actual two-photon absorption will occur in that case. Single-photon loss can also affect the fidelity of the output.

The effects of photon loss can be reduced by heralding on those events in which both photons emerge from the wave guides. This can be accomplished by using a dual-rail encoding and measuring the output of the device, accepting only those events in which both photons emerge in the correct paths. A heralded approach of that kind is scalable when implementing a fusion gate [15] or a CNOT gate based on teleportation [16], since the operation of both types of gates includes a measurement of the output. Postselection of that kind can compensate for photon loss inside the wave guide as well as photons scattered into free space.

In principle, postselection could also be performed by measuring the final state of the atoms and only accepting those events in which they are found in their ground state, but that would be difficult to implement and it would not compensate for loss in the wave guide itself.

## III. Model and simulation

The system used to model a quantum Zeno gate is illustrated by the energy level diagram in Fig. 2. With a maximum of one photon incident in each wave guide, there can be up to two photons in the waveguides labelled $A$ and $B$, as indicated by the blue lines. The coupling of the photons from one wave guide to the other due to their evanescent fields is represented by the dashed arrow with a coupling coefficient of $C$.

For simplicity, we will assume that the photons in each wave guide are coupled to a single three-level atom as indicated by the red lines in Fig. 2, with a continuous coupling between the photons and the atoms. Including many atoms in the model would increase the strength of the two-photon absorption while giving similar results. A Zeno gate could also be implemented using a set of ring resonators containing a single atom, as discussed in Section VII, which is also consistent with the model shown in Fig. 2.

The upper atomic levels are assumed to be resonant with the energy of two photons, while the intermediate atomic state is detuned from single-photon resonance by a frequency of $\Delta$. The interaction Hamiltonian is given in the dipole approximation by $\hat{H}' = -\hat{\mathbf{d}} \cdot \hat{\mathbf{E}}$, where $\hat{\mathbf{d}}$ is the dipole moment of the atomic transition and $\hat{\mathbf{E}}$ is the second-quantized electric field operator. The matrix element of $\hat{H}'$ for the absorption of a photon accompanied by an atomic transition from the ground state to the intermediate state will be denoted by $M$, as illustrated by the dashed arrows in Fig. 2. For simplicity, the matrix element for a transition from the intermediate atomic state to the upper level will be assumed to have the same value. Virtual transitions from the ground state to the upper excited state (via the intermediate state) produce the required two-photon absorption. This is included directly in the Hamiltonian, rather than using a fixed rate of two-photon absorption.

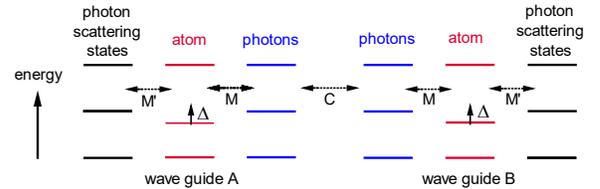

**Fig. 2:** System used to model a quantum Zeno gate. The blue lines represent the presence of up to two photons in wave guides $A$ and $B$, while the red lines represent the states of three-level atoms to produce strong two-photon absorption. Photon loss due to scattering is taken into account by including additional photon scattering states indicated by the black lines. The coupling coefficient C between the two wave guides is represented by a dashed arrow, as are the matrix elements $M$ and $M'$ that couple the incident photons to the atoms and the atoms to the scattered photons, respectively.

An excited atom could emit a photon into a different mode. This corresponds to a photon scattering process that results in the loss of a photon from the system. Losses of this kind were modelled by including an additional photon scattering state for each waveguide that is coupled to the excited atomic states, as indicated by the black lines in Fig. 2. When coupled to a resonant cavity, the use of a single photon scattering state is an excellent approximation, as discussed in more detail in Section VII. In the case of coupled wave guides without a cavity, there would be a continuum of possible scattering states, and this approximation provides a qualitative way to include the effects of photon loss in the model. The matrix element for a transition from an excited atomic state to one of the scattering states will be denoted $M'$, as indicated by the dashed arrows in Fig. 1.

With a maximum of two photons incident in the wave guides, the rotating-wave approximation gives a total of 28 possible states of the system. This corresponds to the usual situation in which photon absorption only occurs in conjunction with atomic excitation. We will denote the photon creation operators in wave guides $A$ and $B$ by $\hat{a}^\dagger$ and $\hat{b}^\dagger$, while the photon creation operators in the corresponding scattering states will be denoted by $\hat{c}^\dagger$ and $\hat{d}^\dagger$. The raising and lower operators for the two atoms will be denoted by $\hat{A}_\pm$ and $\hat{B}_\pm$.

Given that there are only two modes of the electromagnetic field included in the model of Fig. 2, the interaction Hamiltonian $\hat{H}' = -\hat{\mathbf{d}} \cdot \hat{\mathbf{E}}$ in the dipole approximation reduces to

$$
\begin{aligned}
\hat{H}' = {} & C(t)\left[\hat{a}^\dagger \hat{b} + \hat{b}^\dagger \hat{a}\right] \\
& + M(t)\left[\hat{a}^\dagger \hat{A}_- + \hat{a}\hat{A}_+ + \hat{b}^\dagger \hat{B}_- + \hat{b}\hat{B}_+\right] \\
& + M'(t)\left[\hat{c}^\dagger \hat{A}_- + \hat{c}\hat{A}_+ + \hat{d}^\dagger \hat{B}_- + \hat{d}\hat{B}_+\right].
\end{aligned}
\tag{1}
$$

Here $M(t)$ and $M'(t)$ are the time dependent matrix elements of $\hat{H}'$ evaluated between the relevant states of the system. Their time dependence is due to the motion of the photon wave packets. The total Hamiltonian also contains the usual expressions for the energies of the states shown in Fig. 2.

The interaction Hamiltonian of Eq. (1) is a three-level generalization of the widely-used Jaynes-Cummings model for two-level atoms. Two photon absorption occurs as a result of virtual transitions from the ground state of an atom to its upper excited state via the intermediate atomic state.

The photons were assumed to propagate as localized wave packets through the wave guides. As a result, the matrix element $M(t)$ for the coupling to the atoms will be a function of the time $t$ as will the coupling coefficient $C(t)$ between the two wave guides. If these coefficients are turned on and off sufficiently slowly, any probability amplitude for an atom to be left in its intermediate state at the end of the process will vanish as a result of the adiabatic theorem. Thus $M(t)$ and $C(t)$ will be assumed to vary slowly as illustrated in Fig. 3, with maximum values $M_{\max}$ and

$C_{\max}$. In practice, there will be limitations on how slowly these coefficients can be varied and this will be a potential source of error that can be reduced using heralding.

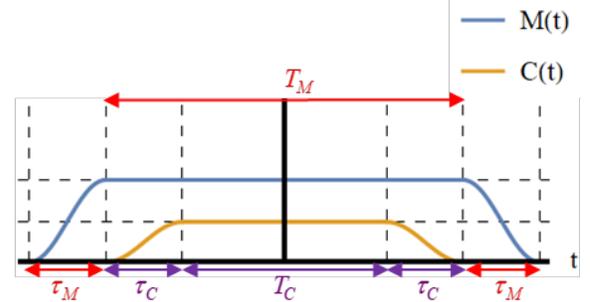

**Fig. 3:** Typical time dependence assumed for the coupling parameters $M(t)$ and $C(t)$ (arbitrary units). The coupling between the wave guides is smoothly turned on and off over a time interval of $\tau_C$ and left at a constant value for a time interval of $T_C$. The corresponding time intervals for the coupling between the photons and the atoms are denoted $\tau_M$ and $T_M$. $C(t)$ was assumed to be nonzero only when $M(t)$ is at its maximum value.

A single photon incident in wave guide $A$ with no photon in wave guide $B$ will have undergone a linear phase shift denoted $\phi_a$ when it emerges from the device. Similarly, $\phi_b$ will denote the linear phase shift experienced by a single photon incident in wave guide $B$. When a photon is incident simultaneously in both wave guides, the system will undergo a different overall phase shift $\phi_{ab}$ due to the Zeno effect as discussed above. We will define the nonlinear phase shift $\Delta\phi_N$ as $\Delta\phi_N = \phi_{ab} - (\phi_a + \phi_b)$.

Under ideal conditions, a single photon entering the device will emerge in the opposite wave guide. It will be convenient to simply exchange the two wave guides (or relabel them) so that an incident photon will emerge in the same wave guide instead. We will also assume that linear phase shifts of $-\phi_a$ and $-\phi_b$ are applied after the photons emerge so that the only net phase shift is the nonlinear phase shift of $\Delta\phi_N$ that occurs when a photon is present in both inputs. Under ideal conditions, the device in Fig. 1 will thus implement a unitary transformation given by

$$
\hat{U} = \begin{bmatrix}
1 & 0 & 0 & 0 \\
0 & 1 & 0 & 0 \\
0 & 0 & 1 & 0 \\
0 & 0 & 0 & e^{i\Delta\phi_N}
\end{bmatrix}.
\tag{2}
$$

Here we have used a basis corresponding to the logical states $|0,0\rangle$, $|0,1\rangle$, $|1,0\rangle$, and $|1,1\rangle$. We will choose the coupling parameters to implement a controlled sign gate with $\Delta\phi_N = \pi$.

An analytic solution to Schrodinger's equation is not feasible when the coefficients in the interaction Hamiltonian of Eq. (1) are time dependent. Schrodinger's equation was therefore solved numerically using Mathematica. The probability amplitudes for the various final states and their associated phases were calculated and used to compute the fidelity of the output state with the desired unitary transformation of Eq. (2). For simplicity, we considered the average fidelity given by

$$F = \overline{\langle \Psi_{in} | \hat{U}^\dagger \hat{\rho}_{out} \hat{U} | \Psi_{in} \rangle}. \qquad (3)$$

Here $|\Psi_{in}\rangle$ represents the input state and $\hat{\rho}_{out}$ is the density matrix of the actual output state. The overbar denotes an average over the four input states in which each qubit has a logical value of 0 or 1. A more general definition could be used but Eq. (3) is sufficient to illustrate the effects of interest.

The calculations for a specific set of parameters each required roughly an hour on a personal computer when the coupling coefficients were varied slowly enough for the adiabatic theorem to be well satisfied.

Departures from the desired output state can occur as a result of several factors, including a two-photon absorption coefficient that is too small as well as photon losses of various kinds. The magnitude of these errors can be reduced using heralding techniques that will be described in more detail in the following sections.

## IV. Errors due to limited two-photon absorption

The operation of the quantum Zeno gate depicted in Fig. 1 relies on strong two-photon absorption. In this section, we analyze the errors that can occur if the two-photon absorption coefficient is too weak. We will assume for now that there is negligible loss due to photon scattering, which corresponds to $M' = 0$. We will also assume that the parameters $M(t)$ and $C(t)$ are turned on and off very slowly so that the adiabatic theorem holds.

Fig. 4 shows the performance of a quantum Zeno gate under these ideal conditions as a function of the matrix element $M_{max}$. The red dots show the performance of the device after heralding on those events in which the atoms are all found in their ground state at the end of the process. The blue dots show the fidelity of the logic operation without any heralding. There is only one point in which there was any significant difference between the two, and the remaining results are shown in red.

It can be seen from Fig. 4(a) that the fidelity $F$ of a quantum Zeno gate will be degraded if the coupling between a photon and the atoms is too small, which corresponds to weak two-photon absorption. It can be seen from Fig. 4(b) that this results in an incorrect value of the nonlinear phase shift. The presence of two photons in the same wave guide will also result in a significant excitation of the upper atomic levels, as reflected in the reduced probability of success for the heralding process in Fig. 4(c).

The angular frequency of the photons was chosen to have the value $\omega = 1$ in arbitrary units with $\hbar = c = 1$. Here $\hbar$ is Planck's constant divided by $2\pi$ and $c$ is the speed of light. The other parameters used in Fig. 4 correspond to $M' = 0$, $\Delta = 0.25$, $C_{max} = 0.00012$, $t_C = 1000$, and $\tau_M = 1000$. $T_C$ and $T_M$ were adjusted along with $M_{max}$ such that a single photon incident in one wave guide would be completely coupled into the other wave guide. This is necessary because the effective coupling between the wave guides is somewhat dependent on the value of $M_{max}$.

Dimensionless parameters were used to simplify the plots and the discussion of the results. The actual values of the parameters will vary from one experiment to another, but the relative magnitude of the parameters is what is most relevant, such as the ratio of the detuning to the coupling strength. Typical parameters for rubidium atoms, for example, can be found in Ref. [23].

Fig. 4 shows that these errors can be all be suppressed in the limit of strong photon-atom coupling and there is no need for heralding in this ideal case. The equivalent effect could be achieved by using a larger number of atoms with a smaller matrix element. In either case, increasing the strength of the photon-atom interaction will also produce larger amounts of photon loss, which can be minimized using heralding as shown in the next section.

Earlier studies of heralded quantum Zeno gates used a model that assumed that photon loss could be described by fixed rates of single-photon loss and two-photon absorption [15,16,18]. In that model, the nonlinear phase shift does not depend on the strength of the two-photon absorption as it does in Fig. 1(b). This can be understood from the fact that a nonlinear medium, such as a collection of three-level atoms, will typically produce a nonlinear phase shift (Kerr effect) in addition to two-photon absorption.

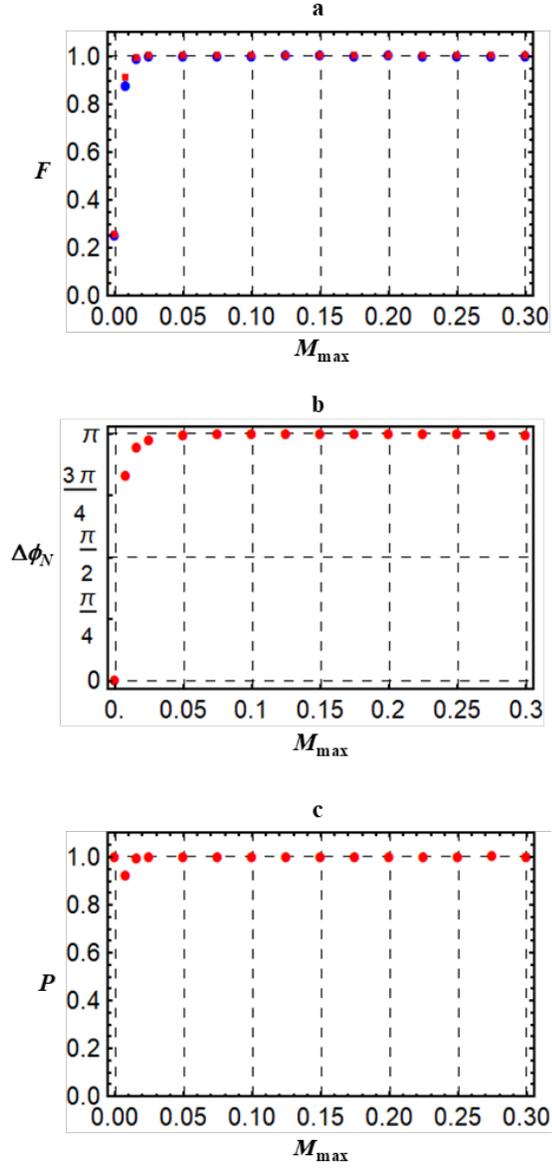

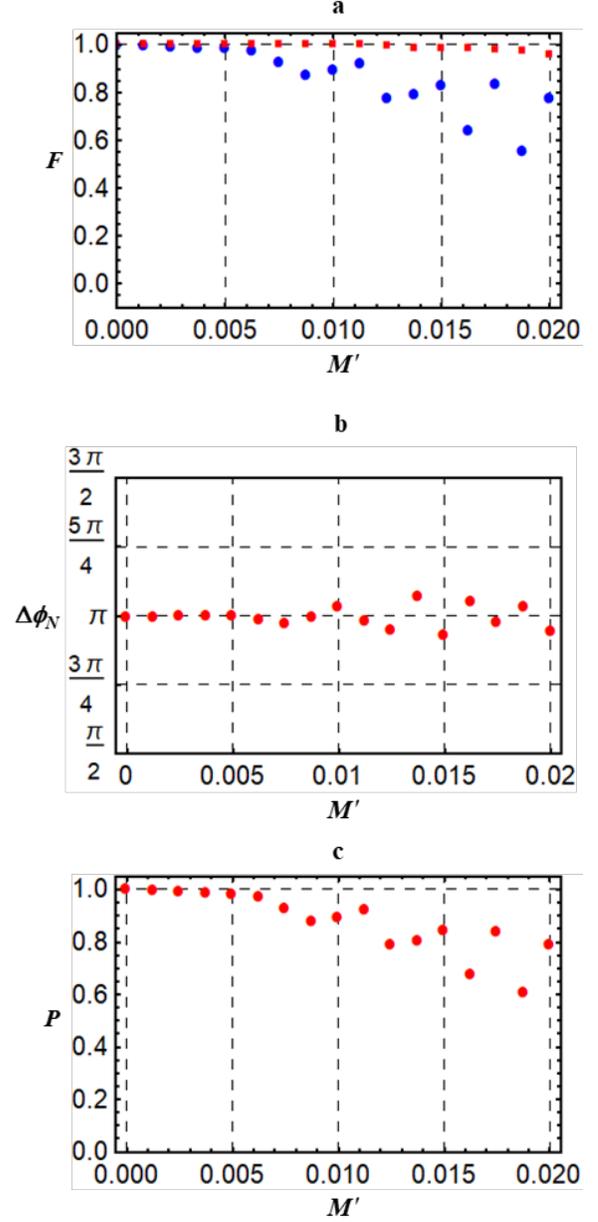

**Fig. 4:** Performance of a quantum Zeno gate as a function of the matrix element $M_{max}$ that couples the photons to the atoms and is responsible for two-photon absorption. Here other potential error sources such as photon scattering have been neglected. (a) The gate fidelity $F$ as a function of $M_{max}$. The results obtained without heralding are indicated by the blue dots, which overlap with the heralded results shown in red except for the initial point. (b) Nonlinear phase shift $\Delta\phi_N$. (c) Probability $P$ of success for the heralding process, in which the results of the logic gate were only accepted if both photons emerge in the correct path.

**Fig. 5:** Performance of a quantum Zeno gate as a function of the photon scattering matrix element $M'$. (a) Gate fidelity $F$. (b) Nonlinear phase shift $\Delta\phi_N$. The heralding process does not correct the value of $\Delta\phi_N$ and the results with and without heralding are both represented by the red dots. (c) Success probability $P$ for the heralding process. It can be seen that heralding can greatly improve the performance of a quantum Zeno gate in the presence of moderate photon scattering.

## V. Photon scattering

The effects of photon scattering are included in our model by the addition of the scattering states shown in Fig. 1. These states could also represent the effects of atomic transitions into other degenerate atomic states that do not couple back into the original photon modes. The rate at which an excited atom will emit photons into these states is dependent on the value of the matrix element $M'$.

The performance of a quantum Zeno gate is plotted in Fig. 5 as a function of $M'$. The results without heralding are shown again in blue while the results with heralding are shown in red. All of the parameters are the same as in the idealized case of Fig. 4 except that $M_{max}$ was fixed at a value of 0.25 while $M'$ was varied. As mentioned previously, heralding of this kind can be used for scalable quantum computing when applied to a fusion gate [15] or a CNOT gate based on teleportation [16].

It can be seen from Fig. 5(a) that the gate fidelity without heralding begins to degrade for sufficiently large values of $M'$. The heralding process will eliminate events in which a photon was scattered, but the fact that the system is coupled to those states can still produce changes in the nonlinear phase as can be seen in Fig. 5(b). The heralding process does not provide any correction to the nonlinear phase shift. Nevertheless, it can be seen that the heralded fidelity is always substantially better than the results without heralding.

The fidelity without heralding shows an increasing amount of oscillatory behavior at larger values of $M'$. A similar effect can be seen in the nonlinear phase shift plotted in Fig. 5(b) as well as the probability of success shown in Fig. 5(c). These oscillations are due to a flow of probability amplitudes between the various states that are on resonance with each other, which is somewhat analogous to the Rabi oscillations of an atom driven by a laser beam.

These results show that heralding can produce a substantial improvement in the fidelity of a quantum Zeno gate in the presence of photon scattering, especially when the scattering rate is moderate. The performance can be further improved using distillation or photon loss codes [15,16,18].

## VI. Errors due to nonadiabatic coupling

The results shown in Figs. 4 and 5 were based on the assumption that the photon-atom matrix element $M(t)$ and the wave guide coupling coefficient $C(t)$ were both turned on and off sufficiently slowly that the adiabatic theorem was well satisfied, as illustrated in Fig. 3. In this section, we will consider the errors that can arise if those parameters are varied sufficiently fast that the adiabatic theorem no longer applies.

Fig. 6 illustrates a situation in which the matrix element $M(t)$ was turned off over a relatively short time interval $\tau_M = 1$ (in arbitrary units), which is three orders of magnitude faster than was used for the results in Figs. 4 and 5. The fidelity $F$ of a quantum Zeno gate is plotted in Fig. 7(a) as a function of $M_{max}$, where the other parameters are the same as in Fig. 4. A comparison of Figs. 4 and 7 show that a rapid change in the coupling between the photons and the atoms can cause a large decrease in the fidelity that can be corrected very well by the heralding process.

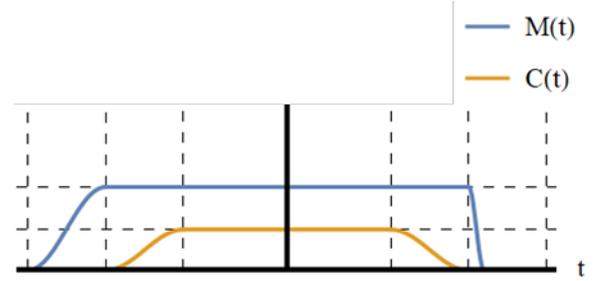

**Fig. 6:** An example of a situation where the coupling $M(t)$ between the photons and the atoms was turned off sufficiently fast that the adiabatic theorem no longer applies. In this example the coupling $C(t)$ between the two wave guides was still turned on and off adiabatically.

Fig. 8 shows the performance of a quantum Zeno gate as a function of $\tau_M$ where the value of $M_{max}$ was held fixed at a value of 0.25. Once again, it can be seen that the heralding process is effective in reducing the errors due to nonadiabatic coupling. The effects of the nonadiabatic coupling is equivalent to having a single-photon loss rate that is larger than would otherwise be expected. From the adiabatic theorem, the error rate due to nonadiabatic coupling (without compensation) depends primarily on the magnitude of the detuning.

## VII. Experimental considerations

The high-fidelity operation of a quantum Zeno gate requires large values of the photon-atom matrix element $M_{max} = -\mathbf{d} \cdot \mathbf{E}$, as can be seen in Fig. 4. The value of $M_{max}$ can be increased by using optical cavities with a small mode volume, since the confinement of a photon to a small volume increases its electric field. The Purcell effect in a cavity also decreases the rate of spontaneous emission by excited atomic states. The use of cavities with small mode volumes is an important aspect of most cavity quantum electrodynamics experiments [7,9,10,11,13].

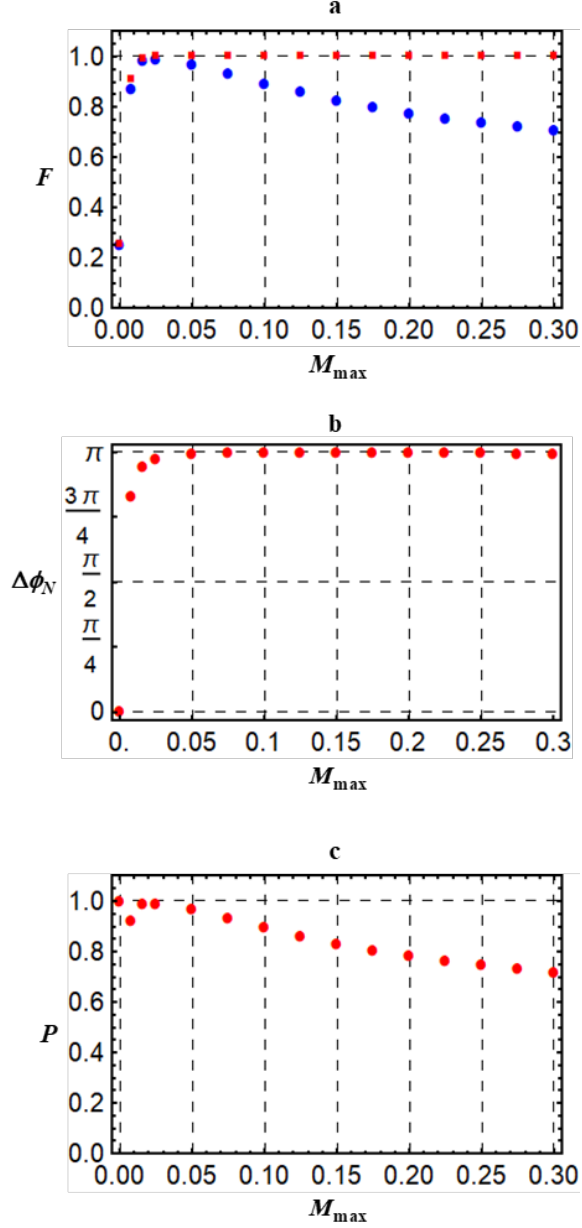

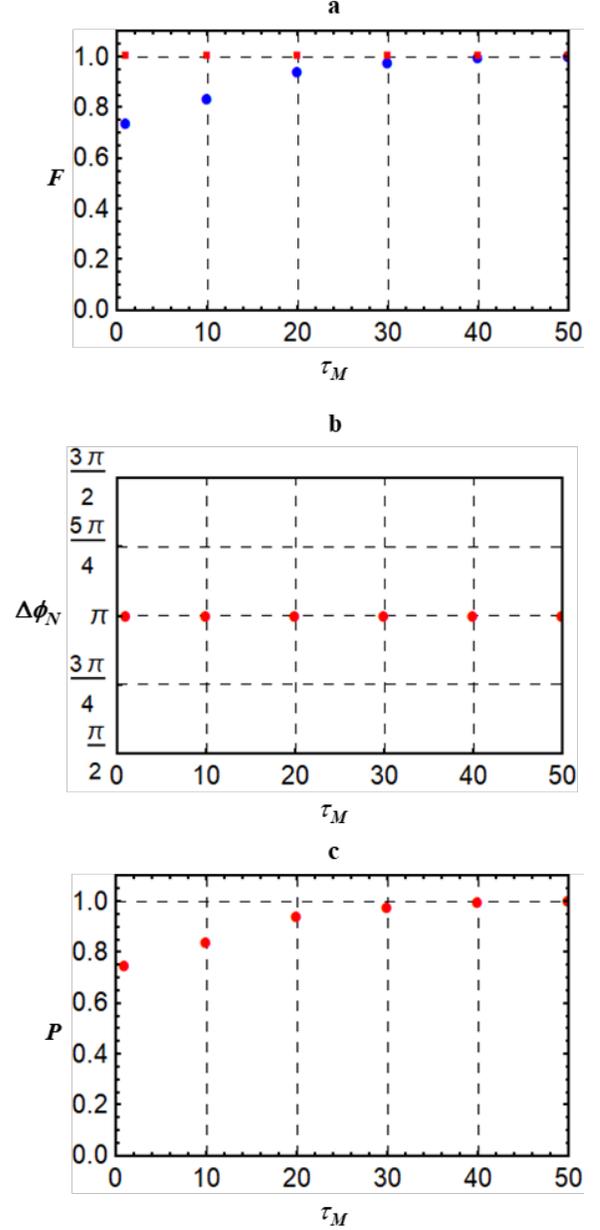

**Fig. 7:** Performance of a quantum Zeno gate under conditions where the matrix element $M(t)$ that couples the photons and the atoms was turned off sufficiently quickly that the adiabatic theorem no longer applies. (a) Gate fidelity $F$. (b) Nonlinear phase shift $\Delta\phi_N$. (c) Success probability $P$ for the heralding process. Once again, the blue points in (a) correspond to the operation without heralding while the red points show the heralded results. It can be seen that heralding is very effective in maintaining a high fidelity under these conditions.

**Fig. 8:** A plot of the performance of a quantum Zeno gate as a function of the time $\tau_M$ (arbitrary units) in which the coupling between the photons and the atoms was turned off. The variables $F$, $\Delta\phi_N$, and $P$ are the same as in Fig. 7. It can be seen once again that the heralding process is very effective in correcting errors of this kind.

A Zeno logic gate equivalent to that shown in Fig. 1 can be implemented using two ring resonators as shown in Fig. 9 [17]. Here the coupled wave guides of Fig. 1 have been replaced by two coupled ring resonators. The resonant cavities could be fabricated using circular wave guides on a substrate or the whispering gallery modes of two microspheres.

Extremely small mode volumes can be obtained using photonic crystal cavities [11].

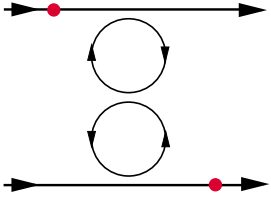

**Fig. 9:** Alternative implementation of a Zeno logic gate. Instead of coupling two long wave guides as in Fig. 1, here two ring resonators are coupled together by their evanescent fields. Two other wave guides serve as the input and output ports. When the coupling coefficients are adjusted properly, a single photon (red dot) incident in one input will be coupled into the other output. The addition of two-photon absorbing atoms will introduce a nonlinear phase shift of $\pi$ as before.

The model shown in Fig. 2 includes a single photon scattering state for each input path. There is a well-known phenomenon in which a photon travelling around the ring resonator is backscattered into a mode in which it travels in the opposite direction due to imperfections in the resonator. Since this cavity mode is on resonance with the original photon and the other cavity modes are not, the assumption of a single scattering mode is an excellent approximation for the ring resonator implementation shown in Fig. 9.

It has been shown that a quantum Zeno gate without heralding would be expected to have performance characteristics similar to those of the more commonly used logic gates based on cavity quantum electrodynamics [23]. Both types of devices have a similar dependence on the cooperativity factor. One advantage of using a Zeno gate is that the coupling strength only needs to be large, which avoids the errors that can occur in conventional cavity QED gates due to variations in the coupling fields.

The results shown in Fig. 5 suggest that the use of heralding could provide an order of magnitude improvement in the value of $(1 - F)$ for quantum Zeno gates. Further experimental investigations would be required to determine the achievable values of the relevant parameters such as $M$ and $M'$. It has been estimated that a quantum Zeno gate should be scalable for quantum computation if the rate of two-photon absorption is approximately a factor of 2200 larger than the single-photon loss rate, which appears to be achievable [15].

## VIII. Summary and conclusions

The performance of a quantum Zeno gate can be greatly improved by applying techniques familiar from linear optics, such as heralding, distillation, and photon loss codes. Previous studies have shown that a combined approach of that kind can outperform linear optics approaches under conditions that appear to be achievable experimentally [15,16,18]. Heralded quantum Zeno gates can be used for scalable quantum computing when used to implement fusion gates [15] and CNOT gates based on teleportation [16], since the output of the gate is always measured in those applications.

Earlier studies of heralded quantum Zeno gates were based on a model that assumed a fixed rate of single-photon loss and two-photon absorption [15,16,18]. In this paper, we have used a different model in which the photons are coupled to three-level atoms that can produce two-photon absorption. Schrodinger's equation was then solved numerically for the combined system of photons and atoms.

Our results show the presence of additional error sources that are not described by the fixed loss model, including errors in the magnitude of the controlled phase shift produced by a Zeno gate as well as additional photon loss due to nonadiabatic coupling. These additional error sources are expected to be most significant when using single atoms coupled to a resonant cavity as illustrated in Fig. 9, which has the potential advantages of a small mode volume and a large Purcell factor.

Errors in the nonlinear phase shift cannot be corrected by heralding alone. It also seems unlikely that they could be corrected in a straightforward way using distillation [15] or photon loss codes [18]. The nonlinear phase errors can be minimized, however, by using sufficiently strong two-photon absorption. These residual errors should be kept in mind when designing quantum Zeno logic gates. Further experimental work will be required in order to determine the achievable parameters and the expected gate fidelity, although earlier studies have suggested that quantum Zeno gates may allow scalable quantum computation under achievable conditions [15,16,18].

### Acknowledgements

We would like to acknowledge many valuable discussions with Todd Pittman. This work was funded in part by the National Science Foundation under grant number PHY-1802472.